\documentclass[a4paper,fleqn,usenatbib]{mnras}

\usepackage{newtxtext,newtxmath}

\usepackage[T1]{fontenc}
\usepackage{ae,aecompl}

\usepackage{graphicx}	
\usepackage{amsmath}	
\usepackage{amssymb}	

\newcommand{\target}{1H~0323+342}

\title[The radio-loud NLS1 1H 0323+342 in a galaxy merger]
{The radio-loud narrow-line Seyfert 1 galaxy 1H 0323+342 in a galaxy merger}

\author[Doi et al.]{
Akihiro Doi,$^{1,2}$\thanks{E-mail: doi.akihiro@jaxa.jp}
Motoki Kino$^{3,4}$,
Nozomu Kawakatu$^{5}$ and  
Kazuhiro Hada$^{6,7}$
\\
$^{1}$The Institute of Space and Astronautical Science, Japan Aerospace Exploration Agency, 3-1-1 Yoshinodai, Chuou-ku, Sagamihara, Kanagawa 252-5210, Japan\\
$^{2}$Department of Space and Astronautical Science, SOKENDAI, 3-1-1 Yoshinodai, Chuou-ku, Sagamihara, Kanagawa 252-5210, Japan\\
$^{3}$Kogakuin University of Technology \& Engineering, Academic Support Center, 2665-1 Nakano, Hachioji, Tokyo 192-0015, Japan\\
$^{4}$National Astronomical Observatory of Japan, 2-21-1 Osawa, Mitaka, Tokyo 181-8588, Japan\\
$^{5}$National Institute of Technology, Kure College, 2-2-11, Agaminami, Kure, Hiroshima 737-8506, Japan\\
$^{6}$Mizusawa VLBI Observatory, National Astronomical Observatory of Japan, 2-12 Hoshigaoka, Mizusawa, Oshu, Iwate 023-0861, Japan\\
$^{7}$Department of Astronomical Science, SOKENDAI, 2-21-1 Osawa, Mitaka, Tokyo 181-8588, Japan
}

\begin{document}

\date{Accepted 2020 May 20. Received 2020 May 19; in original form 2019 June 17}

\pagerange{\pageref{firstpage}--\pageref{lastpage}} 
\pubyear{2020}

\maketitle

\label{firstpage}

\begin{abstract}

The supermassive black holes (SMBHs) of narrow-line Seyfert 1 galaxies (NLS1s) are at the lowest end of mass function of active galactic nuclei~(AGNs) and preferentially reside in late-type host galaxies with pseudobulges, which are thought to be formed by internal secular evolution.   
On the other hand, the population of radio-loud NLS1s presents a challenge for the relativistic jet paradigm that powerful radio jets are exclusively associated with very high mass SMBHs in elliptical hosts, which are built-up through galaxy mergers.  
We investigated distorted radio structures associated with the nearest gamma-ray emitting, radio-loud NLS1 1H 0323+342.  This provides supporting evidence for the merger hypothesis based on the past optical/near-infrared observations of its host galaxy.    
The anomalous radio morphology consists of two different structures, the inner curved structure of currently active jet and the outer linear structure of low-brightness relics.  
Such a coexistence might be indicative of the stage of an established black hole binary with precession before the black holes coalesce in the galaxy merger process.  
1H 0323+342 and other radio-loud NLS1s under galaxy interactions may be extreme objects on the evolutionary path from radio-quiet NLS1s to normal Seyfert galaxies with larger SMBHs in classical bulges through mergers and merger-induced jet phases.   

\end{abstract}


\section{INTRODUCTION}\label{section:introduction}

The scaling relation between the masses of supermassive black holes~(SMBHs) and spheroidal components in galaxies suggests the regulated co-evolution of SMBHs and galaxies \citep[e.g.,][]{Magorrian:1998,Gebhardt:2000,Ferrarese:2000}.  
Elliptical galaxies with very high-mass SMBHs are thought to be the end products through major mergers with the active galactic nucleus~(AGN) feedback mechanism in the quasar era \citep[e.g.,][]{Hopkins:2008,Kormendy:2013}. 
The relation has been established from the samples of elliptical galaxies and disk galaxies with classical bulges hosting SMBHs with relatively large masses ($\ga 10^7 M_{\sun}$).  
On the other hand, narrow-line Seyfert~1 (NLS1) galaxies are a class of AGNs at the lowest end of the SMBH mass function in the local universe \citep[$\la 10^7 M_{\sun}$; e.g.,][]{Peterson:2011,Woo:2015}.  
The samples of active and inactive galaxies with low-mass SMBHs leads to a significant deviation from the linear relation between the black hole and spheroid stellar masses \citep[e.g.,][]{Kormendy:2013,Graham:2015,Davis:2019}.   
This is suggestive of a different evolutionary track in the low-mass regime. 
NLS1 engines predominantly reside in late-type galaxies \citep{Deo:2006} with pseudo-bulges \citep{Orban-de-Xivry:2011,Mathur:2011}.  
Optical observations of host morphology suggests the paucity of the signature of galaxy interactions \citep{Ryan:2007,Ohta:2007}, in addition to the prevalence of strongly barred spirals/disks \citep{Crenshaw:2003,Deo:2006,Ohta:2007}.  
These characteristics are indications of the internal secular evolution, rather than major mergers, for the growing processes of NLS1's central black holes and host galaxies \citep{Kormendy:2004}.

Meanwhile, the radio jet activity potentially has a strong link with the co-evolution of SMBHs and galaxies.  
The relativistic-jet paradigm is known in which radio galaxies and blazars are associated exclusively with elliptical host galaxies harboring very high mass SMBHs \citep{Kotilainen:1998,Kotilainen:1998a,Laor:2000,Sikora:2007}, with only several exceptions \citep{Ledlow:1998,Keel:2006,Hota:2011,Bagchi:2014,Mao:2015,Singh:2015}.  
Additionally, a big fraction of the host galaxies of radio-loud AGNs are associated with recent or ongoing merger events ($92\%^{-14\%}_{+8\%}$ at $z > 1$, \citealt{Chiaberge:2015}; 93\% at $z < 0.2$ and 95\% at $0.2 \leq z < 0.7$, \citealt{Ramos-Almeida:2012}), suggesting a strong link with the relativistic jet triggering mechanism.    
Most of the NLS1s ($\sim93$\%) are radio-quiet\footnote{Radio-loudness $R$ is defined as the ratio of the 5~GHz to the optical {\rm B}-band fluxes \citep{Kellermann:1989}.  A source is considered radio-loud if $R> 10$.} ($R<10$; \citealt{Zhou:2006,Komossa:2006}), indicating poor jet activities as a class.  
Therefore, the hypothesis that the SMBHs of radio-quiet NLS1s are growing in the internal secular process without merger events could be understood as a population out of the framework of the relativistic jet paradigm + merger-triggered jet activity.

On the other hand, radio-loud objects do exist in the NLS1 population.  Observational evidence that the jet activity of the radio-loud NLS1s is associated with host morphology and merger events is still under exploration.      
Since most of radio-loud NLS1s are distant objects, optical/near-infrared observations for investigating their host morphology are relatively difficult.  
There have been some reports for $\gamma$-ray emitting NLS1s: PKS~1502+036 possibly in an elliptical \citep{DAmmando:2018}, FBQS~J1644+2619 possibly in an elliptical \citep{DAmmando:2017} or a barred lenticular galaxy with a pseudobulge possibly in a minor merger \citep{Olguin-lglesias:2017}, PKS~2004-447 in the host with a pseudobulge \citep{Kotilainen:2016}, and \target~in a host galaxy showing a spiral \citep{Zhou:2007} or ring-like morphology suggestive of a merger \citep{Anton:2008}.  
For the radio-loud (not $\gamma$-ray-detected) NLS1s, spiral hosts with pseudobulge and galaxy-interaction/disturbance indications were found in J111934.01+533518.7 and J161259.83+421940.3 \citep{Jarvela:2018} and an observation suggests that IRAS~20181-2244 is hosted by a late-type galaxy in an interacting system of two galaxies \citep{Berton:2019}.  
Recently, \citet{Olguin-Iglesias:2020} reported the results of deep near-infrared imaging for 29~radio-loud NLS1s, strongly indicating that their hosts are preferentially disc galaxies with the signs of pseudobulges (16 sources) and galaxy interaction (18 sources).  Six radio-loud NLS1s show an offset stellar bulge with respect to the AGN, suggesting distorted morphology due to a galaxy merger.  These properties are suggestive of merger-induced jet activity, but are inconsistent with the relativistic-jet paradigm in which radio-loud AGNs are exclusively in elliptical hosts.   
If NLS1s are young objects eventually evolving into broad-line Seyfert galaxies with larger mass SMBHs \citep{Mathur:2000}, radio-loud NLS1s could be extreme objects on the evolutionary path from radio-quiet NLS1s to broad-line Seyferts \citep{Doi:2012} through merger processes.  
The fueling mechanism driven by minor majors has been proposed for the nuclear activity of broad-line Seyfert galaxies \citep{Taniguchi:1999}.

Radio observations provide an another approach to explore the history of host galaxy's merger by investigating distorted radio morphology that possibly contains traces of the history of jet activity.   
X-shaped radio galaxies usually consist of a pair of Fanaroff--Riley class~II (FR~II)-like radio arms with active lobes (``primary lobes'') and a pair of lower-brightness lobes with no hotspots (``secondary lobes'' or ``wings'') as relics.  
The physical origin of the {\sf X}-shaped radio galaxies is often discussed in the framework of (1)~the spin-flip scenario, (2)~hydrodynamic effects, and (3)~the precession model.    
The spin-flip scenario postulates a change of the jet direction due to a sudden flip of black hole's spin axis through a coalescence with another black hole in the final stage of a galactic merger \citep[e.g.,][]{Merritt:2002}.  
The hydrodynamic effects as an interaction with the surrounding interstellar medium in a merger environment (the backflow diversion model \citep[e.g.,][]{Leahy:1984,Capetti:2002a}; the jet-shell interaction model \citep{Gopal-Krishna:2012}) are considered if a characteristically distorted structure is apparent in secondary lobes.  
The precessing jet in a binary black hole system formed via a merger \citep[e.g.,][]{Begelman:1980,Liu:2007} is postulated in the middle stage of a galactic merger before two black holes coalesce, if a helically curved jet structure is observed.  
For investigating the merger experience of radio-loud NLS1s, the large radio structure that tends to exceed the galaxy scale and the high spatial resolution of radio interferometry could surpass optical observation approaches.

\target\ is the nearest ($z=0.0629$; \citealt{Zhou:2007}) object among known $\gamma$-ray-emitting NLS1s \citep{Abdo:2009,Paliya:2014}.  
Thanks to its proximity, \target\ is one of the best sources that can be investigated by direct imaging of jet activity and host galaxy.  
The host galaxy has been resolved on some level and exhibits a one-armed spiral \citep{Zhou:2007} or ring-like morphology, likely associated with a recent violent dynamical interaction of the host galaxy \citep{Anton:2008,Leon-Tavares:2014,Olguin-Iglesias:2020}.   
The analyses of the bulge profile show a S\'{e}rsic index of $n=0.88/1.24$ at the J/K-band (\citealt{Olguin-Iglesias:2020}; see also \citealt{Leon-Tavares:2014} who reported $n \sim 1.2$ based on their model B), suggesting a pseudobulge \citep[$n<2$;][]{Fisher:2008}.  
The black hole mass is certainly low: $3.4 \times 10^7 M_{\sun}$ based on reverberation mapping \citep{Wang:2016}, which is consistent with estimations based on the X-ray variability ($\la 1$--$4 \times 10^7 M_{\sun}$; \citealt{Yao:2015a,Landt:2017,Pan:2018}) and the single-epoch relation between the line width and luminosity ($\sim 1$--$3 \times 10^7 M_{\sun}$; \citealt{Zhou:2007,Leon-Tavares:2014}), but is an order of magnitude smaller than that given by the black hole--bulge mass relation \citep{Leon-Tavares:2014}.  
The parsec~(pc)-scale radio jet is highly relativistic showing a one-sided structure and and superluminal motions \citep{Wajima:2014,Fuhrmann:2016,Lister:2016,Doi:2018,Hada:2018}.  For kpc-scale radio structure, \citet{Anton:2008} presented two radio images showing a core plus a two-sided structure with unusual morphology.  \citet{Anton:2008} briefly discussed the possibility of merger-induced morphology for the anomalous radio jet structure of \target.  No more detailed studies have been made so far for the kpc-scale radio structure of \target.

In the present paper, for \target\ we report a detailed investigation of radio morphology, indicating the history of changing jet axis.       
We assume a $\Lambda$CDM cosmology with $H_0=70.5$~km~s$^{-1}$~Mpc$^{-1}$, $\Omega_\mathrm{M}=0.27$, and $\Omega_\mathrm{\Lambda}=0.73$.  At the distance to 1H~0323+342, an angular size of 1\arcsec corresponds to 1.2~kpc in the projected distance.

\section{Data and radio images}\label{section:data_radioimage}
We retrieved the Karl G.~Jansky Very Large Array~(VLA\footnote{The VLA is operated by the National Radio Astronomy Observatory, which is a facility of the National Science Foundation operated under cooperative agreement by Associated Universities, Inc.}) archival data with the project codes AM0601, AM577, AP0501.  These data have a variety of array configurations (A-, B-, C-, and CD-array) and frequencies ($1.4$--$43$~GHz).  
The VLA A- and C-array data at 1.4~GHz (AP0501) are (probably) the same as \citet{Anton:2008} previously presented.  More improved images are reported in the present paper.   

\begin{figure}
\includegraphics[width=\linewidth]{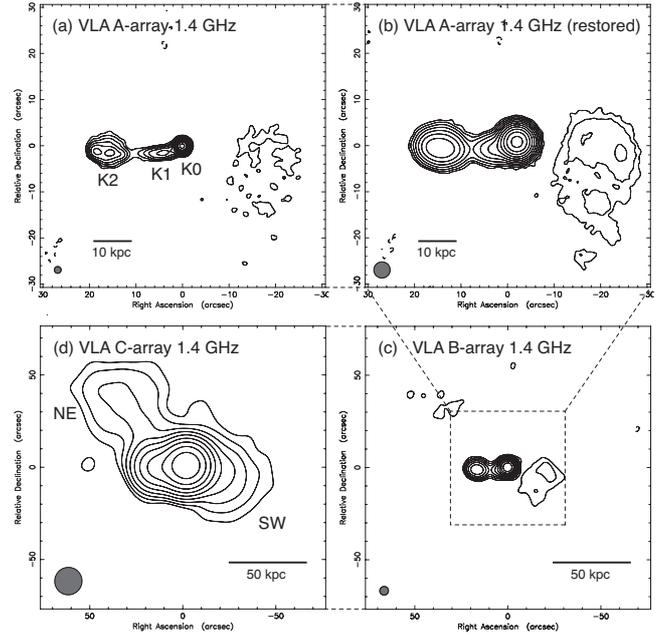}
\caption{Radio images of 1H~0323+342 at kpc scales.  Contour levels are separated by factors of 2 beginning at $3\sigma$ of the rms image noise.  Beam size is illustrated as a gray circle at lower left in each panel.  (a)~~VLA A-array image at 1.4~GHz; (b)~VLA A-array image convolved using a restored beam of $3\farcs5$; (c)~VLA B-array image at 1.4~GHz; (d)~VLA C-array image.  Component names in this paper are indicated in images (a), and (d): K0, K1, and K2 are of a VLA-scale core, a jet, and an inner lobe-like structure at kpc scales, respectively; SW and NE are of a south-west lobe and an north-east lobe as outermost components, respectively.  The image noises are $1\sigma = 0.09, 0.44$, and $0.33$~mJy~beam$^{-1}$ for A-, B-, and C-array images, respectively.  
}
\label{figure:VLAimages}
\end{figure}

Data reduction was performed using the Astronomical Image Processing System~\citep[{\tt AIPS};][]{AIPS:reference} in accordance with the standard procedures for the VLA continuum observation.  Final calibrations were performed by CLEAN deconvolution and self-calibration iteratively using the software {\tt difmap} \citep{DIFMAP:reference}.  We found several radio sources with significant flux contributions in the primary beam of the VLA ($\sim30\arcmin$ at 1.4~GHz) around 1H~0323+342.  We carefully modeled them by the CLEAN procedure for subsequently iterations of self-calibration.  Radio images, in particular at 1.4~GHz, have significantly improved through the rejection of the contaminating emission.  We display final images for each array configuration at 1.4~GHz with natural weighting (Figure~\ref{figure:VLAimages}).  We define the names of components, K0, K1, K2, NE, and SW, as denoted in these images.  

\begin{figure}
\includegraphics[width=\linewidth]{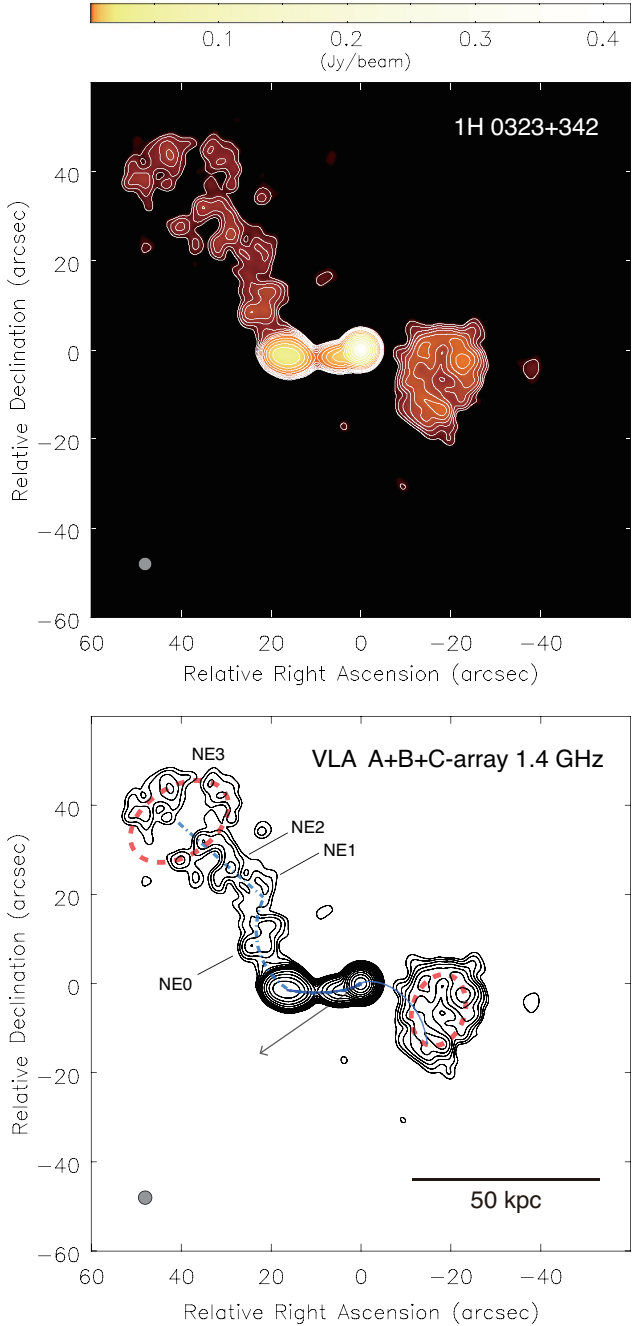}
\caption{Combined VLA A+B+C-array image of 1H~0323+342 with a restoring beam of $3\arcsec$.  Beam size is illustrated as a gray circle at lower left.  Contour levels are $3\sigma \times (-1, 1, \sqrt{2}, 2, 2\sqrt{2}, 4, 4\sqrt{2}, 8, 16, 32,\ldots)$, where $\sigma (=0.053$~mJy~beam$^{-1}$) is the image rms noise.  
({\it Upper}) Radio-intensity map with intensity given by the color bar in units of Jy beam$^{-1}$ and overlaid contours.  ({\it Lower})  Contour map with descriptions/illustrations of radio components and structures.  
Substructures in NE are denoted by NE0--NE3.  The arrow represents the position angle ${\rm PA} = 125 \degr$ for the pc-scale jet.  Red dashed ellipticals are traces of relic lobes (NE3 and SW).  Blue curves represent one of solutions of the recently started precession model; $\beta_{\rm inner}=0.65$, $\beta_{\rm outer}=0$, and $P_{\rm prec}=2.5 \times 10^6$~yr (Section~\ref{section:discussion:precession} and Table~\ref{table:precessionparameter}).  The bold solid, bold dot-dashed, and thin solid curves represent the trajectories of the active jet-lobe, relics, and expected counter jet, respectively.}   
\label{figure:VLAcombined}
\end{figure}

A combined VLA A+B+C-array image at 1.4~GHz was also made (Figure~\ref{figure:VLAcombined}).  After combining all the 1.4-GHz data, each of which was separately preprocessed by self-calibration and subtracting model visibilities of central sources (K0, K1, and K2), we restored visibilities of central components that had been constructed from the A-array data to the combined data, and then made an image via {\tt tclean} in the Common Astronomy Software Applications~(CASA) package \citep{CASA:reference}.  As the result of several trials, we applied Briggs's clean with $\rm{ robust}=0$ \citep{Briggs:1995} and beam restoring of $3\arcsec$.  The NE region seen in the C-array image were resolved into substructures, named NE0--NE3, as denoted in the combined image.         

\begin{figure}
\includegraphics[width=\linewidth]{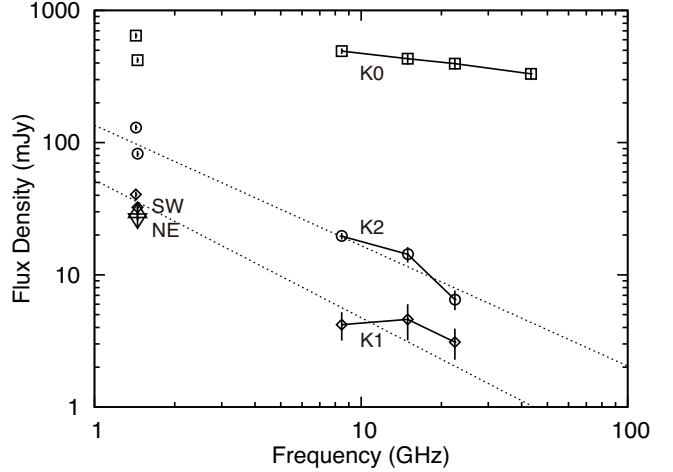}
\caption{Radio continuum spectra of 1H~0323+342.  Squares, diamonds, circles, upward triangle, and downward triangle  represent measured flux densities of K0, K1, K2, SW, and NE, respectively.  Symbols connected with solid lines represent quasi-simultaneous observations.  Dotted lines represent fitted power-law spectra for K1 and K2.}  
\label{figure:spectra}
\end{figure}

The images at the other higher frequencies ($8.4$--$43$~GHz; AM577) are not shown in the this paper, because only a similar radio structure consists of K0, K1, and K2 was observed.  
Figure~\ref{figure:spectra} shows a radio continuum spectrum for each component; the results of flux measurements are listed in Table~\ref{table:data}.   The error of flux density was determined from root-sum-square of the assumption for VLA flux scaling error ($5\%$ at 1.4--8.5~GHz and $10\%$ at 15--43~GHz) and the source identification error in the Gaussian model fitting.

\section{Results}\label{section:results}

\subsection{Radio morphology}\label{section:reslut:radio_morphology}

The armed structure in the northeastern side can easily trace the evolution of jet's position angle from the nucleus to the outermost region, while the emission in the southwestern side exhibits an isolated ring-like structure rather than armed morphology (Figure~\ref{figure:VLAcombined}).  Significant asymmetry with respect to the nucleus is also apparent in terms of the maximum extent and brightness.

In the inner kpc region (Figure~\ref{figure:VLAimages}~(a)), we discovered a well-defined core--jet--lobe structure with edge-brightened morphology that is reminiscent of FR~II radio galaxies.  K0 and K1 are compact and narrow in width, while K2 is significantly resolved in the transverse direction of the jet flow.  The structure elongates eastward with a curve, and reaches $\sim20\arcsec$ ($\sim24$~kpc in the projected distance) and is terminated at K2 in this image.  The one-sided morphology indicates that the K0--K2 structure are in the approaching-jet side.   
K0 is a flat-spectrum core with a spectral index of $\alpha= -0.24 \pm 0.01$ at 8.4--43~GHz, where $\alpha$ is the spectral index in $S_\nu \propto \nu^\alpha$, $S_\nu$ is the flux density at the frequency $\nu$.  K1 and K2 are an optically-thin jet and a lobe showing steep spectra of $\alpha= -1.04 \pm 0.16$ and $-0.91 \pm 0.11$, respectively (Figure~\ref{figure:spectra}).

The northeastern emission~(NE) exhibits an outermost radio emitting region with an elongation up to $\sim60\arcsec$, corresponding to $\sim72$~kpc in the projected distance.  NE was not detected in the VLA a-array images (Figures~\ref{figure:VLAimages}~(a)--(b)), indicating very low brightness.  We define sub-structures NE0--NE3 in NE on Figure~\ref{figure:VLAcombined}; NE3 potentially shows ring-like morphology.  NE1--NE2--NE3 are well-aligned at a position angle of $\sim48\degr$ from the core K0, which is quite different from the position angles of the inner FR~II-like structure.   
NE0 is seen as a bridging emission between the K2 lobe and NE1.  Hence, NE is in the approaching-jet side.  No hotspot-like feature is seen in the NE region.

The southwestern emission~(SW) in low brightness (Figure~\ref{figure:VLAimages}~(d)) has been resolved into an isolated emitting region in high-angular-resolution images (Figures~\ref{figure:VLAimages}~(b) and (c)).  In the combined image (Figure~\ref{figure:VLAcombined}), sub-structures in SW have been revealed to be shell-like morphology.  Therefore, SW and NE3 apparently form a pair of the outermost lobes.  
We cannot find evidence for the centrosymmetric counterpart corresponding to the jet-lobe structure K1--K2.  No bridging emission as a counter jet connecting with SW was detected.  
The total flux density of SW nearly equals to that of NE~(Table~\ref{table:data}).  On the other hand, the asymmetry of the NE--SW morphology is ascribable to the smaller separation distance and higher brightness of SW than those of NE.

Previously, \citet{Anton:2008} presented VLA A- and C-array images at 1.4~GHz, which originated in archival data (probably) the same as we used (AP0501; Figure~\ref{figure:VLAimages} (a) and (d)), and only brief discussions for the radio structure.  Two-sided radio morphology was also pointed out by \citet{Anton:2008} based on their C-array image.  However, NE component was not detected in their A-array image.  The emitting region NE and the internal structures in NE and SW have been first identified in our study.  We point out the FR~II-like morphology of K1--K2 for the first time.  These findings may be ascribable to the improved quality of our images by a careful treatment of contamination sources outside of the field of view (Section~\ref{section:data_radioimage}).

\subsection{The Position-angle profile of radio structures}\label{section:results:PAofjet} 

\begin{figure}
\includegraphics[width=\linewidth]{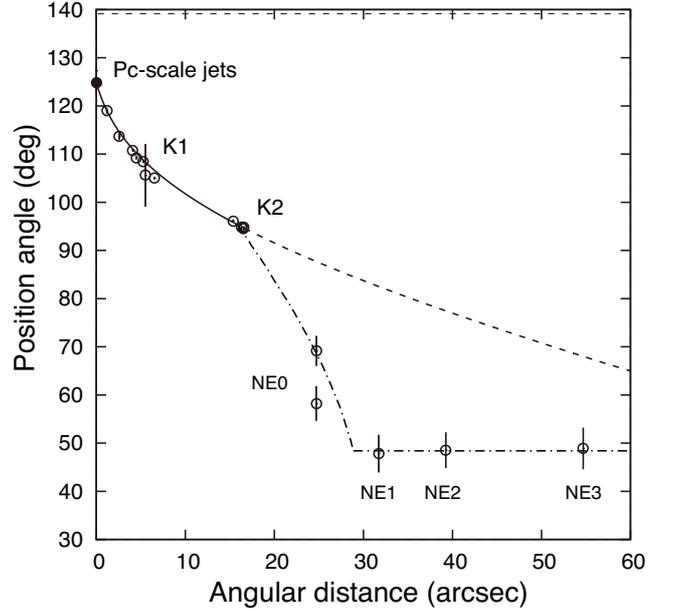}
\caption{Plot of position angle vs.\ angular distance on approaching-jet side.  Open symbols represent measurements on VLA images at 1.4--8.4~GHz.  A filled symbol at ($0\farcs007$, $125\degr$) represents the position angle of pc-scale jets.  Solid and dashed curves represent an expected trajectory by a precession model as an example ($\beta_1=0.7$ and $P_{\rm prec}=2.8\times10^6$~yr; Section~\ref{section:discussion:origin}); note that this example is not a unique solution.  A dot-dashed curve and a line represent an alternative expected trajectory for a precession + deceleration model, which assumes a lower speed ($\beta_2=0.15$) in the outer region as a head speed of the radio lobe.    
}   
\label{figure:positionangle}
\end{figure}

Figure~\ref{figure:positionangle} is the plot tracing the jet structure in position angles with a dependence on the angular distance from the nucleus in the approaching-jet side.   
K1 was reproduced with five sub-components using {\tt modelfit} in {\tt difmap} on Figure~\ref{figure:VLAimages}~(a).  NE0--NE2 were identified using {\tt imfit} in {\tt CASA}.  The position of NE3 is the center of a fitted elliptical.  
The pc-scale jet of 1H~0323+342 is being ejected toward the position angle $PA \sim125 \degr$ in VLBI images \citep{Wajima:2014,Fuhrmann:2016,Doi:2018,Hada:2018}.   
On the other hand, the outermost radio structure NE1--NE3 extends at $PA=48\degr \pm 1\degr$, quite different from that of pc-scale jet.  The NE1--NE3 structure is also well-aligned toward the nucleus.  

The K0-K1-K2 radio structure is curved, and its position angle changes progressively from $PA \sim 120\degr$ to $\sim95\degr$.  The upstream of K1 connects with the pc-scale jet smoothly, while the outermost radio structure NE1--NE3 is not on the extension of the trend of the K0-K1-K2 structure.  NE0 bridges between K2 and NE1.  

It was difficult to trace the distance-dependent structure on the counter side.  The center of the SW lobe is located at $PA \sim -109\degr$, which is slightly misaligned by $\sim21\degr$ with that of NE3 lobe.

\begin{table}
\begin{center}
\caption{Observation data and results of flux density measurements}
\label{table:data}
\begin{tabular}{llrcc}
\hline
\hline
Project	&	Array	&	$\nu$	&	$S_\nu$	&	Comp.	\\
	&		&	(GHz)	&	(mJy)	&		\\
 (1)	&	(2)	&	(3)	&	(4)	&	(5)	\\
\hline
AM0601	&	VLA-B	&	1.43	& $	645.3 \pm 32.3	$ &	K0	\\
	&		&		& $	40.5 \pm 2.1	$ &	K1	\\
	&		&		& $	130.2 \pm 6.5	$ &	K2	\\
AM577	&	VLA-CD	&	8.46	& $	491.7 \pm 24.6	$ &	K0	\\
	&		&		& $	4.2 \pm 0.5	$ &	K1	\\
	&		&		& $	19.6 \pm 1.1	$ &	K2	\\
	&		&	14.96	& $	431.6 \pm 43.2	$ &	K0	\\
	&		&		& $	4.6 \pm 1.4	$ &	K1	\\
	&		&		& $	14.3 \pm 1.9	$ &	K2	\\
	&		&	22.49	& $	396 \pm 39.6	$ &	K0	\\
	&		&		& $	3.1 \pm 0.8	$ &	K1	\\
	&		&		& $	6.5 \pm 1	$ &	K2	\\
	&		&	43.36	& $	330.8 \pm 33.7	$ &	K0	\\
AP0501	&	VLA-A	&	1.45	& $	420.8 \pm 21	$ &	K0	\\
	&		&		& $	32.4 \pm 1.8	$ &	K1	\\
	&		&		& $	82.2 \pm 4.2	$ &	K2	\\
AP0501	&	VLA-C	&	1.45	& $	26.7 \pm 3.5	$ &	NE	\\
	&		&		& $	29.3 \pm 3.6	$ &	SW	\\\hline
\end{tabular}
\end{center}
\begin{footnotesize}
Note. --- Column~1: project code; Column~2: telescope array; Column~3: center frequency; Column~4: flux density; Column~5: component name.  \end{footnotesize}
\end{table}

\section{DISCUSSION}\label{section:discussion}
Key findings of the present study are (1)~a one-sided, FR~II-like jet--lobe structure at the inner kpc-scale emission~(K1-K2), (2)~misalignments in the position angles among the pc-scale jet,  the FR~II-like inner kpc-scale structure with a curve, and an outermost low-brightness radio lobe~(NE), and (3)~a spatially identified low-brightness radio emitting region (SW) on the counter-jet side.  We discuss the possible origin of the apparently peculiar radio morphology in \target.

\subsection{The origin of anomalous radio morphology}\label{section:discussion:origin}

Overall radio morphology of the $\gamma$-NLS1 1H~0323+342 is reminiscent of {\sf X}-shaped radio galaxies \citep[e.g.,][]{Murgia:2001} but exhibits significant asymmetry.  
X-shaped radio galaxies usually consist of a pair of FR~II radio arms with active lobes (``primary lobes'') and a pair of lower-brightness lobes with no hotspots (``secondary lobes'' or ``wings'').  In the case of \target, the inner FR~II-like radio structure K1--K2 is interpreted as a primary lobe, and the outermost emitting region NE is as the secondary lobe.  In this scenario, the primary lobes should be currently energized by the active nucleus.    The inner structure is apparently one-sided, which is consistent with the existence of a jet and an undetectable counter-jet due to Doppler boosting/de-boosting (discussed later in Section~\ref{section:discussion:precession}).  On the other hand, NE exhibit very low-brightness and no hotspots.  It is likely that the NE region is no longer energized by the central engine and a relic as the past jetted activity.  An armed structure is not obvious in the counter-jet side, but the emitting region SW certainly appears in a corresponding opposite range of position angles (between $\sim -80\degr$ and $\sim -135\degr$; Figure~\ref{figure:VLAcombined}).   

The physical origin of the {\sf X}-shaped radio galaxies may be discussed in the framework of (1)~the spin-flip scenario, (2)~hydrodynamic effects, and (3)~the precession model.  The spin-flip scenario postulates a change of the jet direction due to a sudden flip of black hole's spin axis through a coalescence with another black hole in a miner merger \citep[e.g.,][]{Merritt:2002}.  The scenario predicts (i) a rapid change of jet direction, (ii) reorientation only once, and (iii) a stable jet axis after the event.  The spin-flip scenario cannot be applied to the case of \target, because of the observed curved jet structure (K1) up to the active lobe and the presence of transitional structure (NE0) between the position angles of the primary and secondary lobes; the observed structure rather indicates an ongoing gradual reorientation.  
The hydrodynamic effects as an interaction with the surrounding interstellar medium (ISM; the backflow diversion model \citep[e.g.,][]{Leahy:1984,Capetti:2002a} have also been discussed.  In this framework, the jet-shell interaction model \citep{Gopal-Krishna:2012}) have been proposed for cases where the ``{\sf Z}-symmetric'' morphology is apparent in secondary lobes.  For \target, this mechanism is unlikely because of the straight propagation in the outermost structure of NE1--NE3, which is also well-aligned toward the nucleus.  NE1--NE3 are presumably the oldest radio structures, but show no evidence of the experience of hydrodynamical distortion.  

The precession model is consistent with at least the inner part of radio morphology, which exhibits the smoothly curved structure throughout the pc-scale jet, the edge-brightened jet--lobe (K1--K2), and the transitional emission NE0.  The warping instability of an accretion disc \citep{Pringle:1996} would explain a precessing jet direction; however, it is a stochastic rather than a regular precession.  
Alternatively, a precessing jet can also be present in the case of a binary blackhole system formed via a merger \citep[e.g.,][]{Begelman:1980,Liu:2007}.  In the case of \target, previous work has shown that this object hosts a perturbed galaxy, that might have resulted from a merging process \citep{Anton:2008,Leon-Tavares:2014}.   
Hence, we conclude that the model of precessing jet of a binary black hole system is most preferable to explain the origin of the anomalous radio morphology in \target.      

\subsection{Timescales of the galaxy merger and jet activity}\label{section:discussion:timescale}
The optical image of the host galaxy of \target~exhibits colour gradient profiles inside and across the off-centered ring structure that is very similar to those detected in collisional ring galaxies with star formation \citep{Anton:2008}.
The timescales of ring formation ($\sim50$~Myr; \citealt{Mapelli:2012}) and star-forming clumps in interacting and merging systems ($\sim 10$~Myr; \citealt{Hancock:2007}) can be considered for the age as a merger system for \target.

Meanwhile, it takes a considerable amount of time from the start of a galactic merger to the formation of a black hole binary.
The dynamical friction timescale where the captured black hole sinks to the central region of the galaxy is $t_{\rm sink} \sim 2 \times 10^8 {\rm yr}\ (\sigma_{*}/{\rm 200~km~s^{-1}})^5 (m/10^7 M_{\sun})^{-3/4}$, where $\sigma_{*}$ is the stellar velocity dispersion and $m$ is the mass of the captured black hole \citep{Merritt:2002}.  For a black hole mass of $M = 3.4 \times 10^7 M_{\sun}$ for \target~\citep{Wang:2016} $\sigma_{*}\approx120$~km~s$^{-1}$ is expected from an empirical relation $M \approx 0.309 \times (\sigma_*/200~{\rm km~s^{-1}})^{4.38} 10^9 M_{\sun}$ \citep{Kormendy:2013}.  As a result, $t_{\rm sink} \sim 10$--$80$~Myr if we assume $m/M \sim 1/3$--$1/30$ as a typical minor major.  Thus, this estimated timescale is consistent with the merger age of \target.

Such a large time scale for the formation of binary black hole system means that a significantly large radio structure can be formed before the start of jet precession.  
The outermost radio emitting region consisting of NE1--NE3 shows a straight morphology, which is well-aligned toward the nucleus, up to $72$~kpc in projected distance; 
no signature of distortion by jet precession are seen in this outermost structure.  
Assuming a typical advancing speed of $\sim 0.01c$--$0.1c$ for the head of FR~II radio lobes \citep[][and references there in]{Kawakatu:2008}, the estimated age of the radio relic NE3 is roughly $10$~--$100$~Myr (assuming a viewing angle of $\sim10\degr$; Section~\ref{section:discussion:precession}), which is consistent with the possible timescale of the galaxy merger in \target\ as discussed above.  
Hence, the jet activity of \target\ might have started triggered by the galaxy merger.  
A large fraction of radio-loud AGNs are associated with recent or ongoing merger events \citep{Ramos-Almeida:2012,Chiaberge:2015}, which is a strong indication that mergers are the triggering mechanism for the launching of relativistic jets from SMBHs.

On the other hand, the curved structure K1--K2 up to $24$~kpc in the projected distance can be considered as a jet recently formed after the binary system had been established and the precession started.        
The apparent one-sidedness of the K1 jet implies the Doppler beaming effect.  The jet-to-counter jet intensity ratio $R_{\rm I}>69$ (Eq.~\ref{eq:intensityratio_app}, the ratio of a peak intensity of $11$~mJy~beam$^{-1}$ in the K1 jet to three times the image noise $\sigma=0.053$~mJy~beam$^{-1}$ (Figure~\ref{figure:VLAcombined})), constrains the jet speed $\beta > 0.61$ in the units of speed of light.  This leads to an upper limit of the kinematic age of K2, $\la 1$~Myr, if we adopt a viewing angle of $3\degr$ \citep{Abdo:2009}.  

Therefore, the anomalous radio structure of \target\ presumably contains traces of the history of jet activity both before and after the formation of a binary black hole system.  

\subsection{A recently started jet precession model}\label{section:discussion:precession}
In this subsection, we present a toy model of jet precession that started from the formation of a binary SMBH system long after a galaxy merger.  Possible timescales and geometric parameters are provided by this model for the observed radio morphology in \target.

\begin{table*}
\begin{minipage}{135mm}
\begin{center}
\caption{Parameters of recently started jet precession model}
\label{table:precessionparameter}
\begin{tabular}{lccccccccccc}
\hline\hline
Case	&	$\theta_{\rm pc}$	&	$\Omega$	&	$\phi_0$	&	$P_{\rm prec}$	&	$t_{\rm start}$	&	$\theta_{\rm NE1--NE3}$	&	$t_{\rm K2}$	&	$\theta_{\rm K2}$	\\
	&	(deg)	&	(deg)	&	(deg)	&	(yr)	&	(yr)	&	(deg)	&	(yr)	&	(deg)	\\
(1)	&	(2)	&	(3)	&	(4)	&	(5)	&	(6)	&	(7)	&	(8)	&	(9)	\\
\hline																	
$\beta_{\rm inner}=0.65$, $\beta_{\rm outer}=0$	&	3	&	$7.3$	&	$6.4$	&	$2.5\times10^6$	&	$8.8\times10^5$	&	$13$	&	$2.7\times10^5$	&	$7.2$	\\
\hline
\end{tabular}
\end{center}
\begin{flushleft}
\begin{footnotesize}
Note. --- Column~1: the case of jet speeds in the inner and outer kpc scales.  We found good solutions from cases of $0.61 < \beta_{\rm inner} \la 0.7$, $\beta_{\rm outer}\sim0$; Column~2: assumed viewing angle at the pc scale; Column~3: the semi-aperture angle of the precession cone; Column~4: the angle between the precession cone axis and the line of sight; Column~5: the period of precession; Column~6: time from precession start; Column~7: the viewing angle of NE1--NE3; Column~8: the age of K2; Column~9: the viewing angle of K2.  
\end{footnotesize}
\end{flushleft}
\end{minipage}
\end{table*}

We initially attempted to reproduce the profile of the approaching-jet side with a simple precession model \citep{Caproni:2004} that assumes ballistic jets with a constant speed and a constant precession pitch.  The set of parameters are the jet speed $\beta$ in the unit of speed of light, the semi-aperture angle of the precession cone $\Omega$, the angle between the precession cone axis and the line of sight $\phi_0$, and the period of precession $P_{\rm prec}$.  
We searched parameters under constraints of (i)~$\beta>0.61$ and (ii)~the current viewing angle $\theta_{\rm pc} = 3\degr$.  
The first constraint comes from the apparent one-sidedness of the K1 jet (Section~\ref{section:discussion:timescale})   
The second constraint comes from the spectral-energy-distribution modeling including the $\gamma$-ray regime for \target\ \citep{Abdo:2009}; this viewing angle is consistent with the detection of superluminal motion of $\beta_{\rm app}=9.0\pm0.3$ in the pc-scale jet \citep{Lister:2016}.

We obtained good results in the fit to the currently active jet, namely from the pc-scale jet to the outer tip of the K2 lobe (the solid curve in Figure~\ref{figure:positionangle}), 
but, failed in the fit over the entire angular scales with any combinations of parameters (the dashed curve in Figure~\ref{figure:positionangle}).  Note that because of only a limited number of measurements we had to fix the jet speed $\beta$ to determine the rest of precession parameters $P_{\rm prec}$, $\Omega$, and $\phi_0$.  Almost the same trajectories on the sky plane were obtained with various values of $0.61< \beta \la 1$.  However, NE components were largely deviated from the model curves.  
In the cases of $\beta \ga 0.70$ the expected trajectory in the counter-jet side cannot reach the distance of SW region in a single precession cycle, because of huge asymmetry due to the light-travel-time effect (Eq.~(\ref{eq:armlength})).  Hence, we postrate that the jet has become sub-relativistic ($0.61< \beta \la 0.70$) from the highly relativistic ($\beta \sim 0.995$, e.g., \citep{Fuhrmann:2016}) before escaping the pc-scale region.

Next, we modified the precession model by introducing (a)~different advancing speeds in the inner and outer distances and (b)~a start time of the precession.  This stepwise speed profile was intended to switch from a high speed ($\beta_{\rm inner}$) to a low speed ($\beta_{\rm outer}$) at a boundary distance from the nucleus.   
The former modification provided a good result to reproduce the bridging structure of K2--NE0--NE1, when $\beta_{\rm outer} \sim 0$ ($< 0.1$).  This means that the position angle of the shock front changes gradually, while the bulk of accelerated synchrotron electrons stay behind at the same position angle.  This is a similar approach applied to the structure between the active lobe and the tail of wings in {\sf X}-shaped radio galaxies \citep{Gong:2011}.  
By the later modification, before the precession started, we consider that the shock front was continuously supplied with kinetic energy through the jet at $PA = 48\degr$ to form the linear structure that extends further away (NE1--NE3).  Consequently, this recently-started precession model including a deceleration is well-reproduced the NE structure as shown in dot-dashed curve/line in Figures~\ref{figure:VLAcombined} and \ref{figure:positionangle}).   Table~\ref{table:precessionparameter} lists the determined parameters for the case of $\beta_{\rm inner} = 0.65$ as representative case.   For the cases $0.61< \beta \la 0.70$, determined parameters were very similar.

In this scenario, 
the jet speed was $\beta_{\rm inner} \sim 0.6$--$0.7$ in the inner kpc region, and has changed to $\beta_{\rm outer} \sim 0$ at a distance where the outer tip of the K2 lobe is observed.  The jet initially directed to ${\rm PA} \sim 48\degr$ and a viewing angle of $\theta_{\rm NE1--NE3} \sim 13\degr$ and formed the NE1--NE3 structure of linear morphology; a precession started $t_{\rm start} \sim 1$~Myr ago with the period of precession $P_{\rm prec} \sim 3$~Myr, the position angle of the hotspot gradually moved and NE0 was left out as a relic; the kinematic age of the currently observed K2 lobe is $t_{\rm K2} \sim 0.3$~Myr (at the viewing angle $\theta_{\rm K2} \sim 7\degr$) in the observer frame.  The precession continues today as jet's viewing angle decreases ($\theta_{\rm pc} = 3\degr$ at the pc scale) and \target is detected in $\gamma$-rays.     

By using $P_{\rm prec} \sim 600 (r/10^{16}{\rm cm})^{5/2} (M/m) (M/10^8 M_{\sun})^{-3/2}$~yr \citep{Begelman:1980} and assuming a typical minor merger ($m \sim 0.1M$), an orbital radius of $\sim 0.05$~pc is estimated, which is far from the range where the orbit is effectively shrunk via gravitational radiation.  The central engine of \target~may be currently in the intermediate stage of a binary black hole system.  
We tried to detect the expected winding structure in the pc-scale jet in the series of very-long-baseline interferometry~(VLBI) images\footnote{We used VLBI images obtained at 15.3~GHz in the Monitoring of Jets in AGNs with very long baseline array~(VLBA) experiments \citep[MOJAVE;][]{Lister:2005}} of \target\ during our previous study \citep{Doi:2018}.  However, no significant sign beyond measurement errors was detected because the expected amplitude $0.05$~pc corresponding to $0.04$~milli-arcsecond~(mas) was too small in the jet extending to about 10~mas.

\subsection{Relics as a reservoir of past jet powers}\label{section:discussion:relics}
NE0--NE3 seen far from the edge-brightened active lobe K2 was found in very low brightness, and the outermost region NE3 shows no clear evidence of a hotspot.  Hence, NE0--NE3 are no longer energized by the nucleus through jets and is therefore left as radio relics on the approaching-jet side.  

SW exhibits shell-like morphology with low brightness without armed structures, and is likely to form outermost paired lobes with NE3 that also shows a ring-like structure (Figure~\ref{figure:VLAcombined}).  
The northern part of SW ($PA \sim -85\degr$) looks as if it were a counter lobe corresponding to K2, because roughly a comparable separation ($\sim20\arcsec$) was observed on each side.  However, we can rule out the possibility, because the light-travel-time effect (Eq.~(\ref{eq:beta_app})) postulates a much less arm length of $<8\arcsec$ in the counter-jet side in the case of the jet speed $\beta>0.61$ (Eq.~(\ref{eq:armlength})).  
Moreover, the total flux density of SW nearly equals to that of NE, which indicates that the advancing speeds of the both lobes have been decelerated to $\beta \sim 0$ in the framework of Doppler beaming effect (the flux density ratio $R_{\rm F} \sim 1$, Eq.~(\ref{eq:fluxratio_app})).  Hence, the whole emission of SW can be interpreted as a relic on the counter-jet side.

The paired relic of NE/SW exhibits highly asymmetric morphology in terms of the maximum extent, brightness, and position angle.  The arm-length ratio of $R_{\rm L} \sim 60\arcsec/25\arcsec$ (Eq.~(\ref{eq:armlength})) cannot make a compromise with the flux density ratio $R_{\rm F}\sim1$ (Eq.~(\ref{eq:fluxratio_app})) even at any possible values of the jet speed $\beta$ and viewing angle $\theta$.  In addition, significantly higher surface brightness in the counter-jet side~(SW) compared to the approaching side~(NE).  Such properties are possibly caused by jet interactions with an inhomogeneous ISM \citep[e.g.,][]{Gopal-Krishna:2004}.  In the SW region, the same amount of jet kinetic energy may be confined in a cavity smaller than the NE region.  The misalignment by $PA \sim 21\degr$ between the centers of SW's and NE3's elliptical outlines is also ascribable to the interaction in surrounding environments that is asymmetric on approaching- and counter-jet sides.

Using the relation between the radio luminosity and the cavity power in X-ray-emitting hot gas \citep{Cavagnolo:2010}, we estimated the kinetic power of lobes NE and SW to be $\sim10^{43.7}$~ergs~s$^{-1}$, which is comparable with or slightly less than an estimate for the innermost jet that emits $\gamma$-rays \citep[$10^{44}$--$10^{45}$~ergs~s$^{-1}$;][]{Paliya:2014}.  This relatively large power suggests that the past jet activity also had sufficient jet kinetic powers for escaping to kpc scales in the form of supersonic lobes, which make FR~II morphology \citep{Kawakatu:2009}.

Edge-brightened, FR~II-like radio morphology at kpc scales has been evidently discovered in a fraction of radio-loud NLS1s: PKS~0558-504 \citep{Gliozzi:2010}, FBQS~J1644+2619 \citep{Doi:2011a,Doi:2012}, SDSS~J120014.08$-$004638.7 \citealt{Doi:2012}, J0953+2836, J1435+3131, J1722+5654 \citep{Richards:2015}, J0814+5609 \citep{Berton:2018}, SDSS~J103024.95+551622.7 \citep{Rakshit:2018,Gabanyi:2019}, while an FR~I-like one has been found in the radio-quiet/intermediate NLS1 (Mrk~1239, \citealt{Doi:2015}).    
These FR~II-like structures extend up to $\sim10$--$100$~kpc in the projected size, indicating that supersonic jet flows with sufficiently large jet kinetic powers are emanated from the NLS1 central engines even with low masses of SMBHs, in addition to evidence for the long-lasting ($\ga10^7$~yr) FR~II jet activity \citep{Doi:2012}.   
Thus, \target\ in the present study is similar to these radio-loud NLS1s with FR~II-like radio morphology in terms of the jet power and the age of jet activity (Section~\ref{section:discussion:timescale}).  

In these radio-loud NLS1s, significantly curved radio structure on their one side are seen in some cases (FBQS~J1644+2619, J1435+3131, and J0814+5609), although well-aligned paired lobes at opposite directions are mostly observed.  Interestingly, the brightest knot appears at the middle of a radio arm that also exhibits lower-brightness radio emission at the outermost region (J1435+3131 and J0814+5609).  Such an arrangement is similar to the combination of K2 (the inner active lobe) and NE (the outer relic) in \target.  These outermost emissions might be relics abandoned due to the historical change of jet direction by precessions in binary black holes or flips of black hole's spin axis through coalescences in galaxy merger processes.  
Further researches are demanded in the future.  

\subsection{Implications of 1H~0323+342 for the evolution of SMBHs and galaxies}
Our radio observations has provided insight that \target\ is associated with a merging system, from a different approach based on the distorted jet morphology.  The result is supporting the previously reported signs of a recent violent dynamical interaction based on optical/near-infrared observations \citep{Zhou:2007,Anton:2008,Leon-Tavares:2014,Olguin-Iglesias:2020}.   
The anomalous radio morphology, in which inner curved structures of the FR II-like jet and the outer linear structure of relics coexist (Section~\ref{section:results}), is indicative of the stage of an precessing black hole binary before the black holes coalesce in the galaxy merger process on the radio-loud NLS1 \target\ (Sections~\ref{section:discussion:origin} and \ref{section:discussion:timescale}) .

On the other hand, the S\'{e}rsic index based on the surface-brightness analysis for the host galaxy suggests the presence of a pseudobulge (\citealt{Olguin-Iglesias:2020}; see also \citealt{Leon-Tavares:2014}), which is thought to be developed through internal secular evolution with little experience of galaxy mergers.  
Therefore, \target\ is a peculiar AGN, which conflicts with the relativistic-jet paradigm that radio-loud AGNs are exclusively associated with very high mass SMBHs in elliptical hosts, which is thought to be built-up through mergers.

Importantly, \target\ is not the only radio-loud NLS1 associated with a pseudobulge.  \citet{Kotilainen:2016} have discovered a pseudobulge in the host galaxy of the $\gamma$-ray-emitting NLS1 PKS~2004-447.  \citet{Olguin-lglesias:2017} found a barred lenticular morphology with a pseudobulge and a minor merger sign in the $\gamma$-ray-emitting NLS1 FBQS~J1644+2619 \citep[cf.][]{DAmmando:2017}.  \citet{Olguin-Iglesias:2020} reported the signs of disky (pseudo) bulges in many radio-loud NLS1s.  
These examples suggest the powerful relativistic jets can be launched from engines of low-mass SMBHs in pseudobulges, which challenges the conventional paradigm.

Interestingly, the black hole mass of \target\ is $3.4 \times 10^7 M_{\sun}$ based on reverberation mapping \citep{Wang:2016}, which is certainly much lower compared to typical radio-loud AGNs, but at the high-mass end of the NLS1 population.  Similarly, other radio-loud NLS1s also have relatively larger mass black holes \citep{Komossa:2006,Doi:2012}.  
This trend is partly approaching the relativistic-jet paradigm in the sense that relatively large mass black holes are preferentially present in radio-loud objects in the NLS1 population. 
Additionally, unlike radio-quiet NLS1s, the sign of galaxy interaction is frequently associated with radio-loud NLS1s \citep{Olguin-Iglesias:2020} including \target.  
On the hypothesis that a merger triggers jet activity \citep{Chiaberge:2015,Ramos-Almeida:2012}, it is likely that these radio-loud NLS1s became radio-loud from radio-quiet AGNs in the NLS1 population.  
Thus, \target\ and other radio-loud NLS1s in pseudobulges under galaxy interactions might be extreme NLS1s on a way of the evolutionary track to normal Seyferts with larger SMBHs and classical bulges.

\section{Conclusion and Summary}\label{section:conclusion}

The host galaxy of the radio-loud $\gamma$-ray-emitting NLS1 \target\ exhibits a combination of the two contradictory signs, a disturbed morphology due to a recent merger and a pseudobulge suggestive of internal secular evolution with little merger experience before.    
In the present study, we presented detailed investigations of the distorted radio morphology associated with \target.    
The precession may be attributed to the binary black hole system, which were recently formed by capturing the secondary black hole through the merger process.   
The observed peculiar radio morphology is interpreted as the result of changes in the direction of black hole's spin axis, in a framework in which a binary black hole system is formed via a merger process.   
Thus, \target\ is an example among radio-loud NLS1s in pseudobulges under galaxy interactions, which might be extreme NLS1s on a way of the evolutionary track from radio-quiet NLS1s to AGNs with larger SMBHs and classical bulges.      
The conclusion have been made by following summary of our investigations based on radio imaging for \target:   

\begin{itemize}

\item We made radio images at frequencies of 1.4--43~GHz from VLA archival data.  At 1.4~GHz, detailed radio structures of \target\ has been revealed from the combined data obtained with A-, B-, and C-array configurations. 
 
\item The inner kpc-scale jet shows a one-sided, jet--lobe structure terminated at $\sim20\arcsec$ (corresponding to $\sim24$~kpc in projected size) from the nucleus, and exhibits an FR~II-like edge-brightened radio morphology.  This radio structure is curved, smoothly connected with the pc-scale jets, and therefore, currently energized by the central engine.      

\item In the further outside, a low-brightness radio emitting region is distributed linearly up to $\sim60\arcsec$~($\sim70$~kpc in projected size) at a position angle significantly different from those of the inner kpc-scale structure and pc-scale jet.  A low-brightness component in the counter-jet side was also identified, and is separated without the bridging structure from the nucleus.  The two outermost components are almost equal in flux density, indicating substantially decelerated in advancing speeds.  These outer components are left as relics of past jet activities.  

\item The precessing binary black hole scenario in the framework of {\sf X}-shaped radio galaxies is most preferable to explain the anomalous radio morphology.  It potentially took a significant time for dynamical friction to establish the binary black hole system ($\sim 10$--$80$~Myr) from the beginning of the galaxy interaction.  This timescale is comparable to a possible kinematic age of the outermost radio component ($\sim10$--$100$~Myr).  
The jet activity might be triggered by a merger event.  The estimated kinematic age of the inner jet--lobe structure with a curve trajectory is only $\sim 1$~Myr.  Precession may have started when the binary black hole system was established at the galactic center.  

\item We presented a possible solution in a recently-started jet precession model.  The pc-scale and inner kpc-scale jets with a curve (K1--K2) was fitted well with a simple precession model.  
The base of jet is now almost pole-on viewed.  
The low-brightness bridging emission (NE0) can also be reproduced as a relic on the precession trajectory of the shock front.  The outermost low-brightness linear structure (NE1--NE3) is considered as a relic linearly expanded to the initial direction without precession.   

\end{itemize}

\section*{Acknowledgments} 
The Karl G.~Jansky Very Large Array is operated by the National Radio Astronomy Observatory, which is a facility of the National Science Foundation operated under cooperative agreement by Associated Universities, Inc.  This work was partly supported by JSPS KAKENHI Grant Numbers JP18K03656(MK), JP18H03721(MK), JP19K03918(NK), and JP20K04020(AD).

\appendix 
\section{Doppler beaming effect and light-travel-time effect}

If we consider jets of a ballistic motion with a same constant advancing speed on the both side and a same intrinsic intensity/flux density on the both side.    
The apparent arm length ratio ($R_{\rm L}$), 
the observed intensity ratio ($R_{\rm I}$), 
the observed flux density ratio ($R_{\rm F}$) of an approaching and receding jet, and
the apparent jet velocity $\beta_{\rm app}$ are expressed as \citep[e.g.,][]{Ghisellini:1993,Gopal-Krishna:2004}  
\begin{eqnarray}
R_{\rm L}  &=& \frac{1+\beta \cos{\theta}}{1-\beta \cos{\theta}}, 
\label{eq:armlength} \\
R_{\rm I}  &=& \left( \frac{1+\beta \cos{\theta}}{1-\beta \cos{\theta}} \right)^{2-\alpha},\label{eq:intensityratio_app} \\    
R_{\rm F}  &=& \left( \frac{1+\beta \cos{\theta}}{1-\beta \cos{\theta}} \right)^{3-\alpha},\label{eq:fluxratio_app} \\    
\beta_{\rm app} &=& \frac{\beta \sin{\theta}}{1-\beta \cos{\theta}}, \label{eq:beta_app} 
\end{eqnarray}
where $\beta$ is the jet speed in the unit of speed of light, $\theta$ is the viewing angle.

\bibliography{mypaper}

\label{lastpage}
\end{document}